\documentclass{aa} 

\usepackage{graphicx}
\usepackage{txfonts}
\usepackage{lipsum}
\usepackage{subcaption}         
\usepackage{lscape}             
\usepackage{placeins}           

\renewcommand{\linenumbers}{}

\begin{document}

\title{Direct observational evidence that higher-luminosity type 1 active galactic nuclei are most commonly triggered by galaxy mergers}
\titlerunning{Higher-luminosity type 1 active galactic nuclei are more commonly triggered by galaxy mergers}

\author{
Yongmin Yoon\inst{1}\thanks{yyoon@knu.ac.kr}
\and
Yongjung Kim\inst{2,3}\fnmsep\thanks{yongjungkim@sejong.ac.kr}
\and
Dohyeong Kim\inst{4}\fnmsep\thanks{dh.dr2kim@gmail.com}
\and
Kyungwon Chun\inst{5}
\and
Woowon Byun\inst{5}
}
\institute{
Department of Astronomy and Atmospheric Sciences, College of Natural Sciences, Kyungpook National University, Daegu 41566, Republic of Korea
\and
School of Liberal Studies, Sejong University, 209 Neungdong-ro, Gwangjin-Gu, Seoul 05006, Republic of Korea
\and
Department of Physics and Astronomy, Sejong University, 209 Neungdong-ro, Gwangjin-Gu, Seoul 05006, Republic of Korea
\and
Department of Earth Sciences, Pusan National University, Busan 46241, Republic of Korea
\and
Korea Astronomy and Space Science Institute (KASI), 776 Daedeokdae-ro, Yuseong-gu, Daejeon 34055, Republic of Korea
}
\authorrunning{Yoon et al. }

\abstract{
We examine the connection between galaxy mergers and the triggering of active galactic nuclei (AGNs) using a sample of 614 type 1 AGNs at $z<0.07$, along with a control sample of inactive galaxies matched to the AGNs for comparison. We used tidal features, detected in deep images from the DESI Legacy Imaging Survey, as direct evidence of recent mergers. We find that the fraction of type 1 AGN hosts with tidal features ($f_T$) is higher for AGNs with higher luminosities and (to a lesser extent) more massive black holes. Specifically, $f_T$ rapidly increases from $0.05\pm0.03$ to $0.75\pm0.13$ as the luminosity of the [O\,{\footnotesize III}] $\lambda$5007 emission line ($L_\mathrm{[O\,{\footnotesize III}]}$), an indicator for bolometric AGN luminosity, increases in the range $10^{39.5}\lesssim L_\mathrm{[O\,{\footnotesize III}]}/(\mathrm{erg\,s}^{-1}) \lesssim10^{42.5}$. In addition, $f_T$ increases from $0.13\pm0.03$ to $0.43\pm0.09$ as black hole mass ($M_\mathrm{BH}$) increases in the range $10^{6.0}\lesssim M_\mathrm{BH}/M_{\odot}\lesssim10^{8.5}$. The fraction $f_T$ also increases with the Eddington ratio, although the trend is less significant compared to that with $L_\mathrm{[O\,{\footnotesize III}]}$ and $M_\mathrm{BH}$. The excess of $f_T$, defined as the ratio of $f_T$ for AGNs to that of their matched inactive counterparts, exhibits similar trends, primarily increasing with $L_\mathrm{[O\,{\footnotesize III}]}$ and weakly with $M_\mathrm{BH}$. Our results indicate that, in the local Universe, galaxy mergers are the predominant triggering mechanism for high-luminosity AGNs, whereas they play a lesser role in triggering lower-luminosity AGNs. Additionally, strong events, such as galaxy mergers, may be more necessary to activate massive black holes in more massive galaxies due to their lower gas fractions. 
}


\maketitle

\section{Introduction}\label{sec:intro}
 An active galactic nucleus (AGN) is a highly energetic region found at the center of a galaxy.  The energy released by AGNs originates from a supermassive black hole (SMBH) surrounded by an accretion disk of gas and dust. As the material falls toward the SMBH, gravitational and frictional processes heat the disk, producing intense radiation, mainly in the ultraviolet to optical wavelength range. This radiation photoionizes gas on multiple spatial scales. The compact, high-density gas close to the SMBH ($\lesssim$0.1–1 pc; \citealt{Abuter2024}) constitutes the broad-line region, where high bulk motion velocities broaden permitted lines to widths of several $10^3$ km s$^{-1}$. On larger scales (hundreds to thousands of parsecs; \citealt{Schmitt2003}), lower-density ionized gas forms the narrow-line region, which produces permitted and forbidden lines with typical widths of a few times $10^2$ km s$^{-1}$.
 
In type 1 AGNs, emission lines from both the broad- and narrow-line regions are visible, whereas in type 2 AGNs only the narrow emission lines are seen. According to the AGN unified model \citep{Antonucci1993,Urry1995}, the difference between type 1 and type 2 arises from the line-of-sight viewing angle of the AGN structures. Additional emission, such as radio waves, can arise from jets launched by interactions between magnetic fields and the spinning SMBH \citep{Blandford1977,Urry1995}. Even in the absence of such luminous jets, AGNs can emit radio waves through various mechanisms, such as AGN-driven winds and free-free emission from photoionized gas \citep{Panessa2019}.

Active galactic nuclei are efficient tools for probing several key aspects of the Universe. Some AGNs are extremely luminous, with bolometric luminosities exceeding $10^{46}$ erg s$^{-1}$, allowing them to be detected at redshifts higher than 6 and enabling studies of the early Universe \citep{Kim2015a,Kim2019,Kim2020,Jiang2016,Wang2021,Sacchi2022}. Moreover, AGNs are optimum laboratories for examining physics under extreme gravitational and electromagnetic conditions \citep{Dovciak2004,Jovanovic2008,Blandford2019}. In addition, AGNs provide insights into galaxy evolution, as they are expected to play a significant role in regulating star formation and the mass growth of galaxies \citep{DiMatteo2005,Bower2006,Croton2006,Hopkins2008,Arjona2024} through powerful winds and jets. This coevolution between AGNs and their host galaxies is reflected in the remarkably tight correlations between SMBH mass and host galaxy properties \citep{DiMatteo2005,Kormendy2013}.
 
Active galactic nuclei can be triggered through various channels. For instance, the supply of cold gas from hot halos of galaxy clusters, as well as ram pressure within clusters, can provide the necessary fuel for AGN activity \citep{Gaspari2013,Li2014,Tremblay2016,Poggianti2017}. Secular processes, including the influence of bars and disk instabilities, can also drive gas inflows that ignite AGNs \citep{Shlosman1989,Crenshaw2003,Ohta2007,Hirschmann2012}. In addition, galaxy mergers play an important role, as the induced tidal torques cause gas to lose angular momentum and flow toward the galaxy center, where accretion onto the SMBH can then power AGN activity \citep{DiMatteo2005,Springel2005,Hopkins2008,Capelo2015}.

Previous observational studies have attempted to determine whether galaxy mergers can trigger AGNs. Several studies have shown that AGNs are more likely to be found in interacting or post-merger galaxies compared to inactive galaxies (e.g., there is an excess of the merger fraction, suggesting that mergers can trigger AGN activity \citep{Carpineti2012,Cotini2013,Ellison2013,Hong2015,Marian2020,Pierce2022,Araujo2023,Hernandez2023,Li2023,Comerford2024}. Some studies have found that dust-obscured high-luminosity AGNs \citep{Kim2015b,Kim2023,Kim2024a,Kim2024b,KI2018} are associated with galaxy mergers \citep{Urrutia2008,Glikman2015}. However, several studies found no evidence of the connection between mergers and AGNs \citep{Gabor2009,Cisternas2011,Kocevski2012,Sabater2015,Mechtley2016,Villforth2014,Villforth2017,Marian2019,Shah2020,Zhao2022}. Furthermore, it has been found that higher-luminosity AGNs are most likely to be triggered by mergers \citep{Treister2012,Ellison2013,Urbano2019,Pierce2022,Pierce2023,Hernandez2023,Tang2023,Euclid2025}, while other studies did not find such a trend \citep{Villforth2014,Marian2020,Steffen2023,Comerford2024}. Therefore, despite extensive examinations in numerous studies, the connection between galaxy mergers and AGN triggering, as well as its dependence on AGN luminosity, remains a controversial issue that has yet to be fully resolved.

In this study we aim to obtain a more definitive answer by investigating the problem using a large sample of type 1 AGNs and tidal features detected in deep images. While type 2 or other types of AGNs have been thoroughly examined in the previous studies, large samples of type 1 AGNs have not been widely used due to the challenges in sample selection, which requires sophisticated spectral fitting, and due to their bright unobscured AGN emission, which can hinder detailed investigation of the host galaxies. A major strength of using type 1 AGNs lies in their utility for probing essential AGN properties, such as black hole (BH) masses and Eddington ratios. Specifically, probing the Eddington ratio can reveal hidden but important aspects, as it can provide a more balanced perspective on how mergers influence BH activity by identifying low-mass BHs with low AGN luminosities but high Eddington ratios (e.g., \citealt{Greene2007}), which are often blended with regularly accreting massive BHs.

Tidal features, which are remnants of stellar debris from galaxy mergers and can be detected through deep images \citep{Quinn1984,Barnes1988,Hernquist1992,Feldmann2008}, provide a powerful tool for studying the AGN--merger connection. Because they can persist for several gigayears after a merger, tidal features help overcome the limitation that observations capture only a snapshot of the Universe \citep{Schweizer1992,Tal2009,Kaviraj2011,Sheen2012,Sheen2016,Duc2015,YL2020,Yoon2022,Yoon2023,Yoon2024a,Yoon2024b,Bilek2023,Hernandez2023}. Based on merger signatures detected in deep images from the Dark Energy Spectroscopic Instrument (DESI) Legacy Imaging Survey \citep{Dey2019}, we examined whether galaxy mergers are a significant triggering mechanism of AGNs and, if so, which AGN properties are most closely associated with mergers, using a large sample of 614 type 1 AGNs at $z<0.07$. By doing so, we offer a more definitive understanding of the relation between AGNs and galaxy mergers.

In this study, the cosmological parameters are taken to be $H_0=70$ km s$^{-1}$ Mpc$^{-1}$ for the Hubble constant, $\Omega_{\Lambda}=0.7$ for the dark energy density, and $\Omega_m=0.3$ for the matter density. The luminosity of the [O\,{\footnotesize III}] $\lambda$5007 emission line (or the bolometric AGN luminosity) is expressed in units of erg s$^{-1}$, and the BH mass in solar masses ($M_{\odot}$).
\\

\begin{figure}
\includegraphics[width=\linewidth]{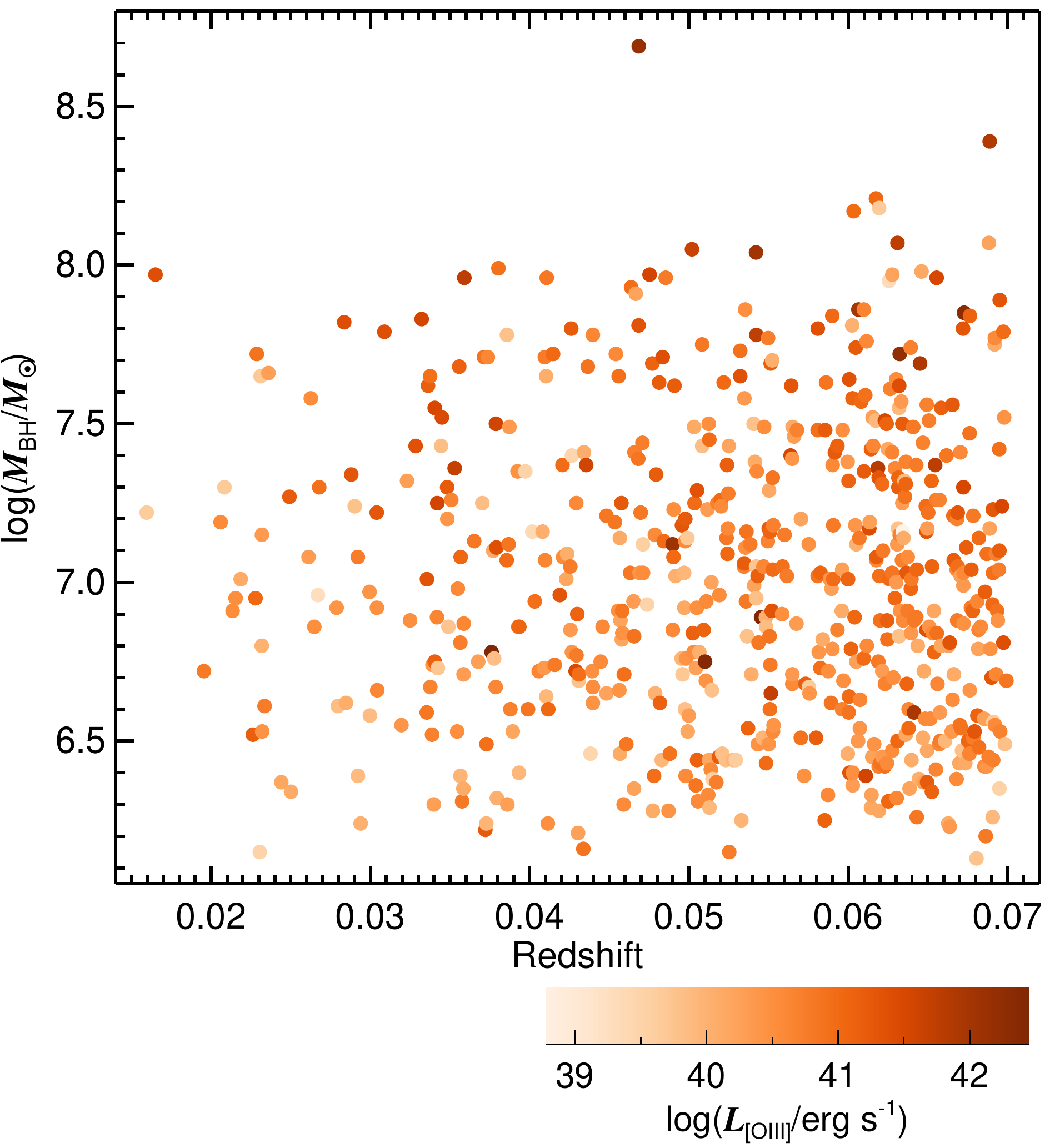}
\centering
\caption{Distributions of type 1 AGNs in the redshift vs. $\log M_\mathrm{BH}$ plane. Type 1 AGNs are indicated by colored circles, with the color representing $\log L_\mathrm{[O\,{\footnotesize III}]}$. See the color bar for the color scale. 
\label{fig:samp}}
\end{figure} 

\begin{figure}
\includegraphics[width=\linewidth]{sampledist.pdf}
\centering
\caption{Histograms of the normalized distributions of redshift and $\log M_\mathrm{BH}$ (or equivalently, $\log M_\mathrm{bul}$) for the type 1 AGN sample and the matched inactive control sample.
\label{fig:match}}
\end{figure}

\begin{table*}[h!]
\caption{Properties of type 1 AGNs.
\label{table}}
\centering
\begin{tabular}{lcccccccc}
\hline\hline
\tiny{SDSS ObjID} & \tiny{R.A.}\tablefootmark{a} & \tiny{decl}.\tablefootmark{a} & \tiny{Redshift} & \tiny{$\log L_\mathrm{[O\,{\footnotesize III}]}$}\tablefootmark{b} & \tiny{$\log L_\mathrm{bol}$}\tablefootmark{c} & \tiny{$\log M_\mathrm{BH}$}\tablefootmark{d} & \tiny{$\log\lambda_\mathrm{Edd}$}\tablefootmark{e} & \tiny{Tidal feature}\tablefootmark{f}\\
\hline
\tiny{587741727654543452} & \tiny{190.38112} &  \tiny{26.04266} & \tiny{0.01596} & \tiny{39.70} & \tiny{43.25} & \tiny{7.22} & \tiny{-2.1} & \tiny{N}\\
\tiny{587739811560882185} & \tiny{214.49812} &  \tiny{25.13687} & \tiny{0.01651} & \tiny{41.43} & \tiny{44.97} & \tiny{7.97} & \tiny{-1.1} & \tiny{Y}\\
\tiny{587725039018311737} & \tiny{180.30980} &  \tiny{-3.67807} & \tiny{0.01955} & \tiny{40.84} & \tiny{44.39} & \tiny{6.72} & \tiny{-0.4} & \tiny{N}\\
\tiny{587733080809668634} & \tiny{171.40070} &  \tiny{54.38255} & \tiny{0.02060} & \tiny{40.62} & \tiny{44.17} & \tiny{7.19} & \tiny{-1.1} & \tiny{Y}\\
\tiny{588017721176686710} & \tiny{169.03189} &  \tiny{41.39811} & \tiny{0.02084} & \tiny{39.72} & \tiny{43.27} & \tiny{7.30} & \tiny{-2.1} & \tiny{N}\\
\tiny{587741489825775725} & \tiny{145.51997} &  \tiny{23.68526} & \tiny{0.02135} & \tiny{40.60} & \tiny{44.15} & \tiny{6.91} & \tiny{-0.9} & \tiny{N}\\
\tiny{587742903405707293} & \tiny{203.02007} &  \tiny{17.04897} & \tiny{0.02154} & \tiny{40.57} & \tiny{44.11} & \tiny{6.95} & \tiny{-0.9} & \tiny{N}\\
\tiny{587732771575955522} & \tiny{144.55110} &  \tiny{7.72779} & \tiny{0.02186} & \tiny{40.16} & \tiny{43.70} & \tiny{7.01} & \tiny{-1.4} & \tiny{N}\\
\tiny{588023669707440191} & \tiny{181.12363} &  \tiny{20.31630} & \tiny{0.02263} & \tiny{41.20} & \tiny{44.75} & \tiny{6.52} &  \tiny{0.1} & \tiny{N}\\
\tiny{587729158970736727} & \tiny{204.56613} &  \tiny{4.54259} & \tiny{0.02279} & \tiny{41.07} & \tiny{44.61} & \tiny{6.95} & \tiny{-0.4} & \tiny{Y}\\
\tiny{587738618103922723} & \tiny{182.68449} &  \tiny{38.33619} & \tiny{0.02287} & \tiny{40.93} & \tiny{44.47} & \tiny{7.72} & \tiny{-1.4} & \tiny{N}\\
\tiny{587742061616037971} & \tiny{200.22362} &  \tiny{21.91952} & \tiny{0.02306} & \tiny{39.56} & \tiny{43.10} & \tiny{6.15} & \tiny{-1.2} & \tiny{N}\\
\tiny{587739407321006093} & \tiny{190.43562} &  \tiny{35.06277} & \tiny{0.02311} & \tiny{39.74} & \tiny{43.29} & \tiny{7.65} & \tiny{-2.5} & \tiny{N}\\
\tiny{587738616483282976} & \tiny{154.95622} &  \tiny{33.36771} & \tiny{0.02318} & \tiny{40.05} & \tiny{43.59} & \tiny{6.80} & \tiny{-1.3} & \tiny{N}\\
\tiny{587726033846272089} & \tiny{150.52933} &  \tiny{3.05769} & \tiny{0.02319} & \tiny{40.40} & \tiny{43.94} & \tiny{7.15} & \tiny{-1.3} & \tiny{N}\\
\hline
\end{tabular}
\tablefoot{The full table is available at the CDS.\\
\tablefoottext{a}{The units of R.A. and decl. are degrees.}
\tablefoottext{b}{The logarithmic luminosity of the [O\,{\footnotesize III}] $\lambda$5007 emission line. $L_\mathrm{[O\,{\footnotesize III}]}$ is expressed in units of erg s$^{-1}$.}
\tablefoottext{c}{The logarithmic bolometric luminosity of the AGN. $L_\mathrm{bol}$ is expressed in units of erg s$^{-1}$ and is converted from $L_\mathrm{[O\,{\footnotesize III}]}$ using the relation $L_\mathrm{bol}\approx3500\,L_\mathrm{[O\,{\footnotesize III}]}$ \citep{Heckman2004}.}
\tablefoottext{d}{The logarithmic BH mass. $M_\mathrm{BH}$ is expressed in solar masses ($M_{\odot}$).}
\tablefoottext{e}{The logarithmic Eddington ratio.}
\tablefoottext{f}{The presence of tidal features in the host galaxy: Y = present, N=absent.}
}
\end{table*}

\begin{figure*}
\includegraphics[scale=0.43]{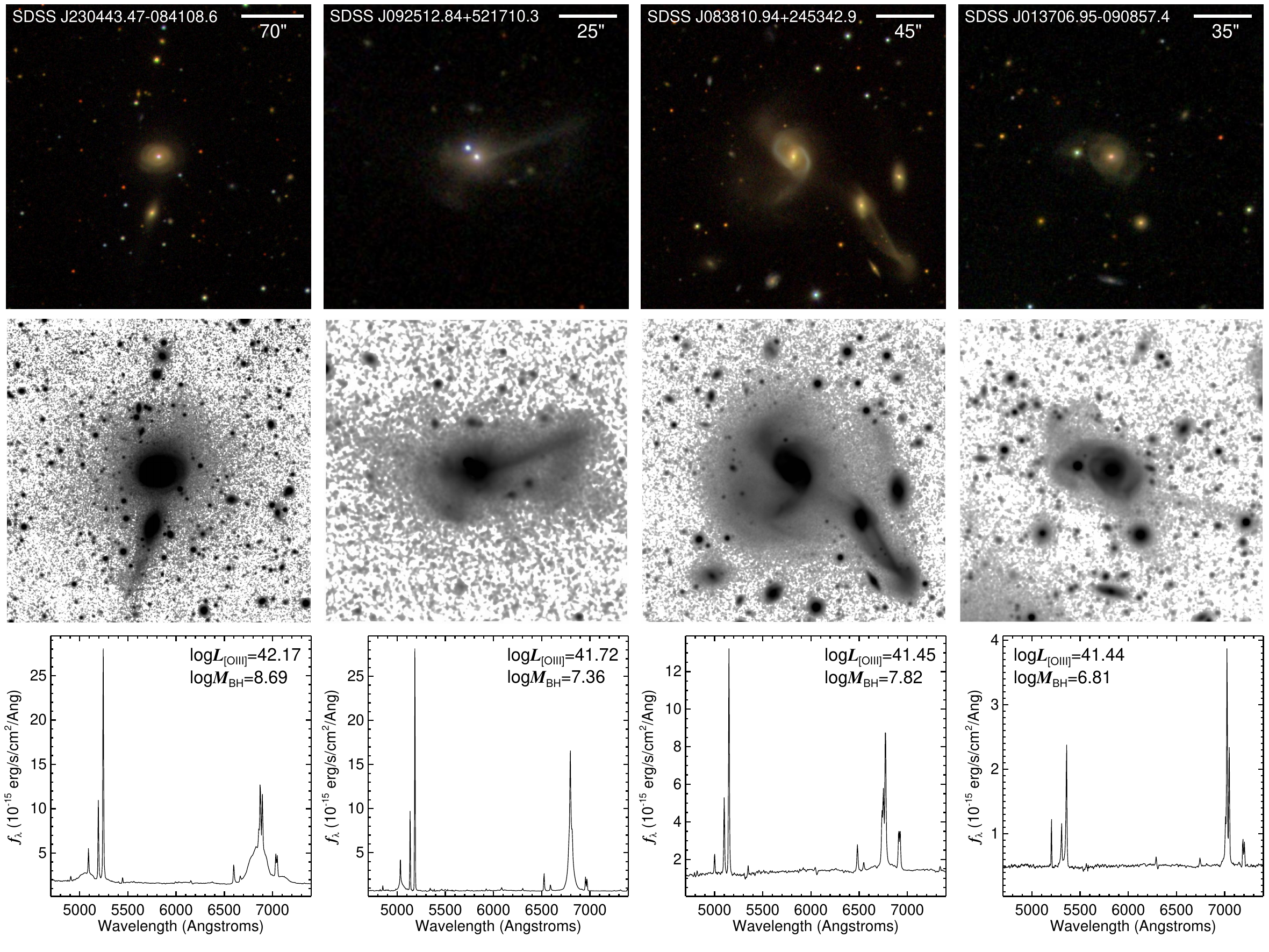}
\centering
\caption{Examples of type 1 AGN hosts with tidal features, which have $\log(L_\mathrm{[O\,{\footnotesize III}]}$/erg s$^{-1})>41.4$. First row: Color images from SDSS. The galaxy ID is provided. The horizontal bar indicates the angular scale of the image. Second row: $r$-band deep images of the DESI Legacy Imaging Survey. The angular scale of the deep image matches that of the color image in the first row. Third row: Optical spectra of type 1 AGNs covering the observed wavelength range 4700-7400\AA, in which H$\beta$, [O\,{\footnotesize III}], and H$\alpha$ emission lines are visible. Also shown are the values of $\log L_\mathrm{[O\,{\footnotesize III}]}$ and $\log M_\mathrm{BH}$, where $L_\mathrm{[O\,{\footnotesize III}]}$ is in units of erg s$^{-1}$ and $M_\mathrm{BH}$ is in units of solar mass ($M_{\odot}$).
\label{fig:ex_1}}
\end{figure*}

\begin{figure*}
\includegraphics[scale=0.43]{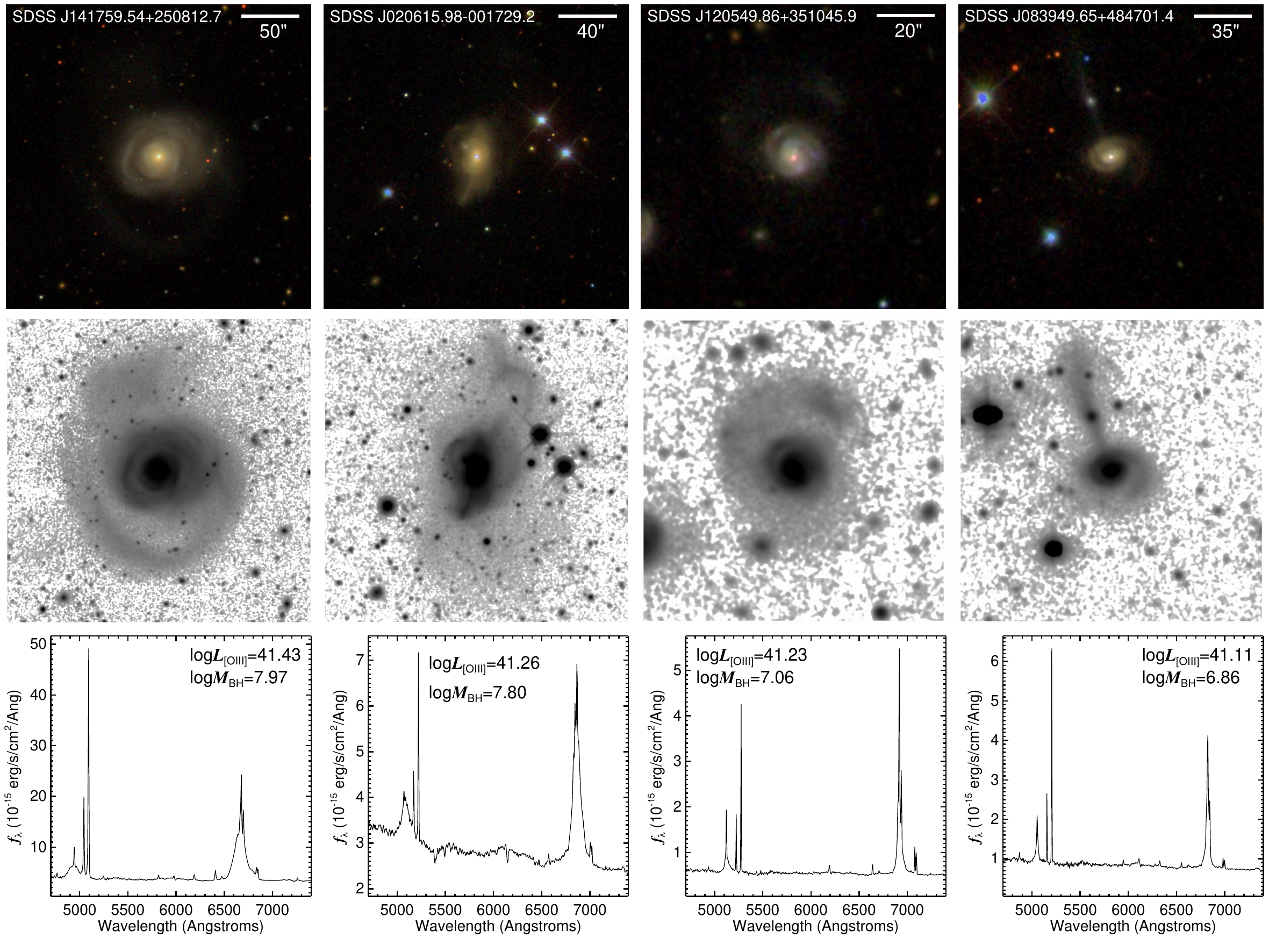}
\centering
\caption{Same as Fig. 3 but for $41.1<\log(L_\mathrm{[O\,{\footnotesize III}]}$/erg s$^{-1})<41.5$. \label{fig:ex_2}}
\end{figure*}

\begin{figure*}
\includegraphics[scale=0.43]{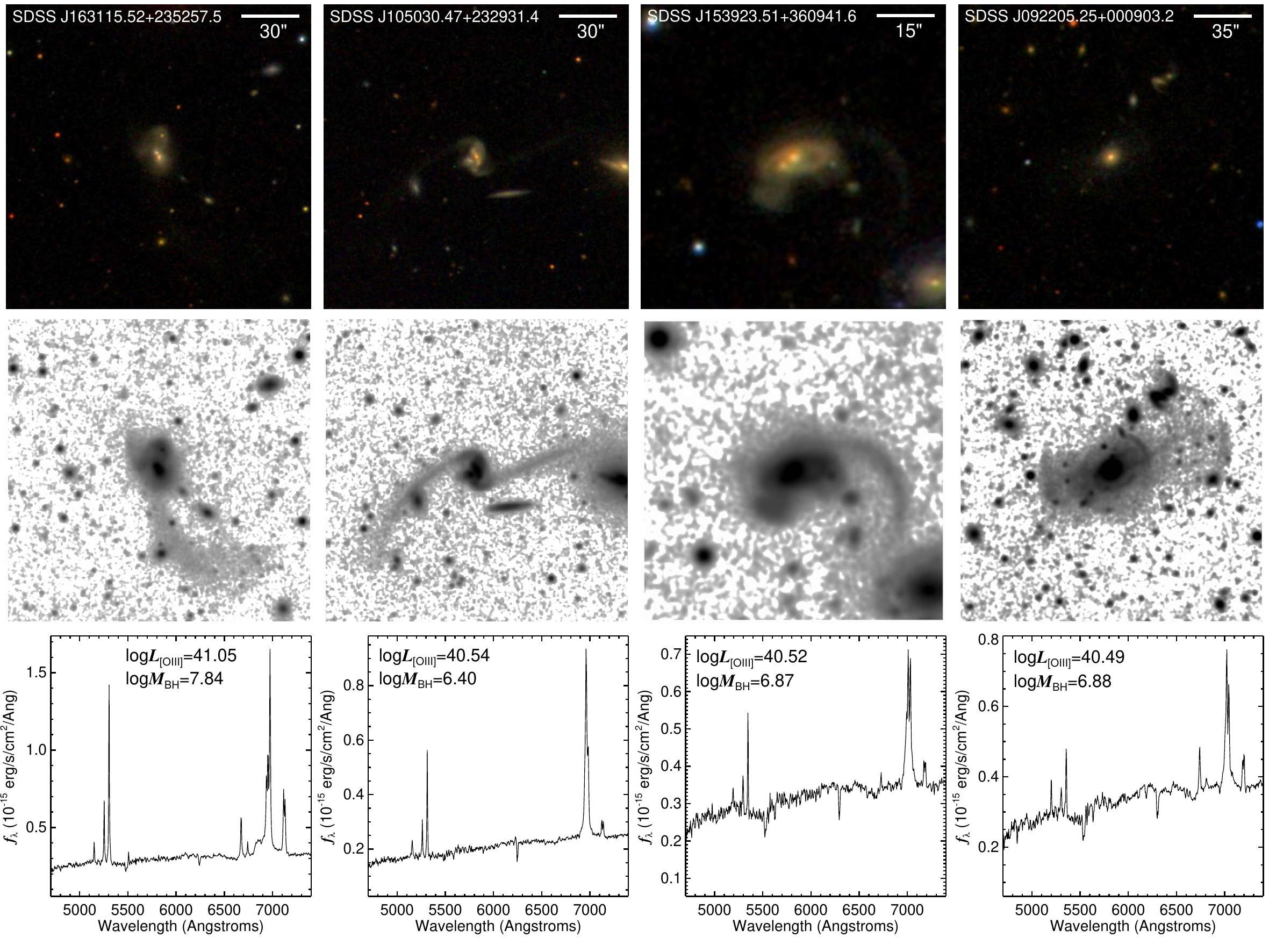}
\centering
\caption{Same as Fig. 3 but for $40.4<\log(L_\mathrm{[O\,{\footnotesize III}]}$/erg s$^{-1})<41.1$. \label{fig:ex_3}}
\end{figure*}

\begin{figure*}
\includegraphics[scale=0.43]{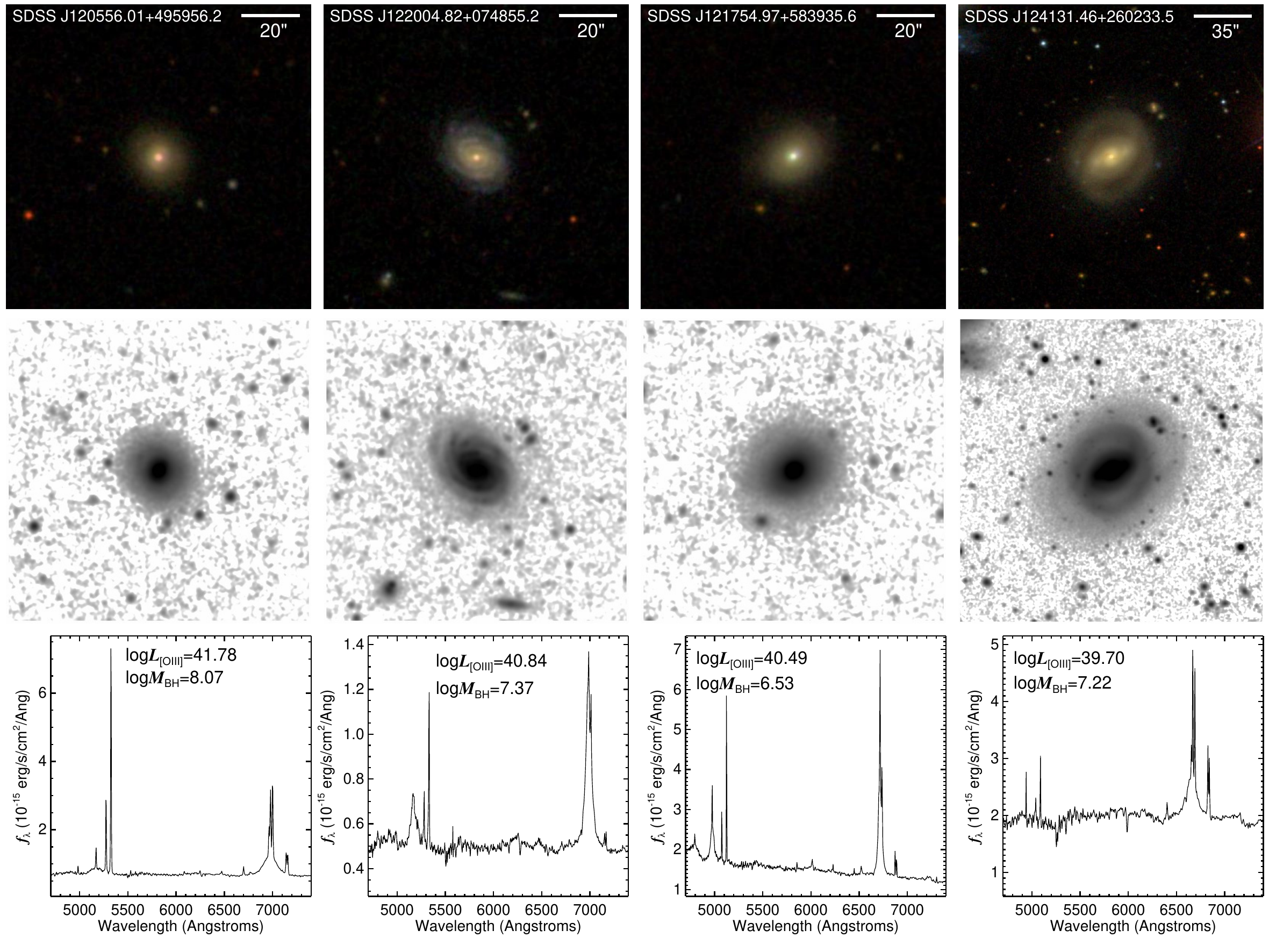}
\centering
\caption{Same as Fig. 3 but for type 1 AGN hosts that do not have tidal features. The AGNs are ordered by decreasing $L_\mathrm{[O\,{\footnotesize III}]}$ from left to right. 
\label{fig:ex_ntf}}
\end{figure*}

\section{Sample}\label{sec:sample}

The sample of AGNs used in this study is from the catalog of type 1 AGNs from \citet{Oh2015}. \citet{Oh2015} cataloged 5553 type 1 AGNs, including low-luminosity AGNs with weak broad-line regions, identified through the analysis of galaxy spectra in the Sloan Digital Sky Survey (SDSS) Data Release (DR) 7. The selection criteria for the type 1 AGNs are as follows: a redshift ($z$) of $<0.2$, a full width at half maximum (FWHM) of broad H$\alpha$ lines exceeding $800$ km s$^{-1}$, and an amplitude-over-noise ratio\footnote{It is defined as the ratio between the amplitude of the Gaussian-modeled emission line and the dispersion of the continuum residual.} of broad H$\alpha$ lines higher than $3$, along with additional criteria such as a completeness cut of $>90\%$ derived from simulations using mock spectra.

In this study we used type 1 AGNs at $z<0.07$. This redshift cut represents a compromise between ensuring a statistically sufficient number of AGNs in the sample and avoiding higher-redshift AGN hosts, which are more likely to be affected by small angular sizes and the cosmological surface brightness dimming effects (see Eq. (6) in \citealt{YP2020}), making tidal feature detection more challenging. The initial sample includes 640 AGNs at $z<0.07$. Of these, 26 AGNs are excluded due to their poor image quality mainly caused by proximity to bright sources (Sect. \ref{sec:tidal}). Thus, the final sample for this study comprises 614 AGNs. We note that our results do not depend on redshift within the range we set ($z<0.07$), which we confirmed by dividing the sample into a couple of redshift bins and applying the same analysis, although the statistical significance is reduced due to the smaller sample sizes. 

Here, the luminosity of the [O\,{\footnotesize III}] $\lambda$5007 emission line ($L_\mathrm{[O\,{\footnotesize III}]}$) is used as an indicator of AGN luminosity, as is commonly adopted in previous studies \citep{Alonso2007,Oh2015,Ellison2013,Comerford2024}.\footnote{Our main findings are similarly obtained when using the luminosity of the broad H$\alpha$ emission as a proxy for AGN luminosity, instead of $L_\mathrm{[O\,{\footnotesize III}]}$. However, to ensure consistency and facilitate comparison with other studies, we adopted $L_\mathrm{[O\,{\footnotesize III}]}$ as an indicator of AGN luminosity throughout this study.} We converted $L_\mathrm{[O\,{\footnotesize III}]}$ into the AGN bolometric luminosity ($L_\mathrm{bol}$) by using the correction relation of $L_\mathrm{bol}\approx3500\,L_\mathrm{[O\,{\footnotesize III}]}$ (with a variance of 0.38 dex) from \citet{Heckman2004}. As in \citet{Oh2015}, we used BH masses ($M_\mathrm{BH}$) derived from the single-epoch mass estimation method developed by \citet{Greene2005}, which is based on the line width and luminosity of the broad H$\alpha$ emission estimated in \citet{Oh2015}. The Eddington ratio ($\lambda_\mathrm{Edd}$) was defined as $L_\mathrm{bol}/L_\mathrm{Edd}$, where $L_\mathrm{Edd}$ is the Eddington luminosity, which is a function of $M_\mathrm{BH}$. To illustrate the general properties of the AGNs, Fig. \ref{fig:samp} displays the distributions of type 1 AGNs in the redshift versus $\log M_\mathrm{BH}$ plane, with the color of the circles indicating $\log L_\mathrm{[O\,{\footnotesize III}]}$ for the AGN sample. In Table \ref{table} we present the properties of type 1 AGNs in our sample, including information on the presence of tidal features in their host galaxies.

The $\log L_\mathrm{[O\,{\footnotesize III}]}$ range of our AGN sample is above $\sim39.5$, as a consequence of the AGN selection criteria used in the catalog of \citet{Oh2015}. Together with other criteria, \citet{Oh2015} selected AGNs based on the signal strength of the broad H$\alpha$ line and a completeness cut, which is mainly determined by the broad H$\alpha$ line luminosity. Such criteria can result in a lower limit on $L_\mathrm{[O\,{\footnotesize III}]}$, given its strong correlation with broad H$\alpha$ line luminosity (Spearman's rank correlation coefficient $=0.6$).

For a quantitative comparison, we defined inactive control sample based on galaxies in SDSS. Previous studies commonly matched control sample based on redshift and mass \citep{Marian2020,Araujo2023,Li2023,Tang2023,Avirett2024,Byrne2024}. According to \citet{McApine2020}, including additional matching parameters beyond redshift and mass is not likely to be crucial, as the recovered results are not sensitive to the choice of matching criteria. In this study, the inactive galaxies in the control sample are matched to type 1 AGNs based on $M_\mathrm{BH}$ and redshift. For each AGN, we matched two inactive galaxies that either lack emission lines or exhibit only emission lines classified as star-forming\footnote{The emission line classification information is obtained from the Max Planck Institute for Astrophysics--Johns Hopkins University (MPA--JHU) catalog (\url{http://www.sdss.org/dr17/spectro/galaxy_mpajhu/}).} according to the criterion of the Baldwin, Phillips, and Terlevich diagram \citep{Baldwin1981}, as defined by \citet{Kauffmann2003} and \citet{Brinchmann2004}. 

Applying the BH scaling relation between $M_\mathrm{BH}$ and $M_\mathrm{bul}$ presented by \citet{Kormendy2013},\footnote{$M_\mathrm{BH}/10^9\,M_{\odot}=0.49(M_\mathrm{bul}/10^{11}\,M_{\odot})^{1.17}$} the $M_\mathrm{BH}$ values of inactive galaxies are derived from their bulge stellar masses ($M_\mathrm{bul}$), which are obtained from the \citet{Mendel2014} catalog. This catalog provides bulge, disk, and total stellar mass estimates based on the two-dimensional bulge+disk decompositions\footnote{The decompositions are conducted using the de Vaucouleurs plus exponential disk model.} performed by \citet{Simard2011}. AGNs and inactive galaxies are matched within a redshift range of $0.002$ and a $M_\mathrm{BH}$ range of $\sim0.1$ dex. The number of inactive galaxies in the control sample is 1228.

In Fig. \ref{fig:match} we compare the normalized distributions of redshift (left panel) and $\log M_\mathrm{BH}$ (right panel) for the type 1 AGN sample and their matched inactive control sample. The two samples show nearly identical distributions, as intended through the matching process.

 \citet{Osborne2024} evaluated the uncertainties and systematic errors in the bulge-to-total light ratios (based on the de Vaucouleurs plus exponential disk model) from \citet{Simard2011} catalog by comparing them with values derived from higher-resolution \textit{Hubble} Space Telescope images. They reported a standard deviation of 0.2 and a median offset of 0.04 in the bulge-to-total light ratios, which implies that $M_\mathrm{bul}$ can be constrained with acceptable accuracy using SDSS images, despite their fundamental limitations. We note that the fraction of inactive galaxies with tidal features marginally increases with $M_\mathrm{bul}$ from $0.07\pm0.01$ to $0.10\pm0.02$ across the full 1.5 dex range of $M_\mathrm{bul}$ in our sample. Thus, the uncertainties and systematic errors in the estimation of $M_\mathrm{bul}$ have little effect on our main results. 
\\

\section{Detection of tidal features}\label{sec:tidal}
We used images from the DESI Legacy Survey DR10 \citep{Dey2019} to detect tidal features. The DESI Legacy Survey comprises three wide-area surveys: the Dark Energy Camera Legacy Survey, the Beijing--Arizona Sky Survey, and the Mayall $z$-band Legacy Survey. Together, these surveys cover a total area of approximately 14,000 square degrees. The $g$- and $r$-band images from the DESI Legacy Survey reach a median surface brightness limit of about 27 mag arcsec$^{-2}$, as defined by the $1\sigma$ background noise within a $1\arcsec\times1\arcsec$ area. This surface brightness limit is similar to that of the deep co-added $r$-band images of the Stripe 82 region of SDSS \citep{YL2020,Yoon2022,Yoon2023}, which are commonly used to identify low surface brightness tidal features around galaxies (e.g., \citealt{Kaviraj2010,Schawinski2010,Hong2015}).

Tidal features are identified through visual inspection of the $g$- and $r$-band images,\footnote{The image depth in the $g$ and $r$ bands is approximately 1 magnitude deeper than that in the $z$ band.} along with composite color images that combine the $g$, $r$, and $z$ bands. During the visual inspection of the images, we individually adjusted pixel value scales (e.g., contrast stretching) and applied Gaussian smoothing with different kernel sizes for all AGN hosts and inactive galaxies in order to enhance faint signals and better identify diffuse tidal features.

Tidal features typically take the form of streams, tails, and multiple shells \citep{Duc2015,Mancillas2019,Bilek2020,Bilek2023,Sola2022}. Tidal streams are thin, elongated features that often appear as fine filaments and are typically linked to minor mergers. In certain cases, these streams are directly connected to smaller companion galaxies. Tidal tails are broad, stretched features that clearly protrude from the host galaxies. These elongated stellar structures, which can form during major mergers, have a similar appearance to tidal streams but are wider and can even extend to the scale of the host galaxies. In some instances, however, tails and streams are not easily distinguishable from one another \citep{Bilek2020,Sola2022}. Tidal shells have the form of arc-like features with distinct, sharp boundaries. These arcs can either follow a common alignment or be randomly distributed around the host galaxy. Shells located farther from the galaxy tend to be more diffuse. Some galaxies display a combination of different types of tidal features. In this study, tidal tails, streams, and shells, around galaxies are collectively categorized as tidal features.\footnote{For instance, the AGN hosts in the first column of Fig. \ref{fig:ex_1} and in the fourth column of Fig. \ref{fig:ex_2} exhibit tidal streams. The AGN hosts in the first and third column of Fig. \ref{fig:ex_2} show tidal tails. The AGN host in the fourth column of Fig. \ref{fig:ex_3} has tidal shells.}

We find that 109 out of 614 type 1 AGN hosts have tidal features ($17.8\%$). Examples of SDSS color images and deep $r$-band DESI Legacy Survey images of type 1 AGN hosts with tidal features are shown in the first and second rows of Figs. \ref{fig:ex_1}--\ref{fig:ex_3}, while those of type 1 AGN hosts that do not have tidal features are displayed in the first and second rows of Fig. \ref{fig:ex_ntf}. The third rows of these figures display optical spectra of type 1 AGNs covering the observed wavelength range 4700-7400\AA, in which H$\beta$, [O\,{\footnotesize III}], and H$\alpha$ emission lines are visible. We also discover that 88 out of 1228 inactive galaxies in the control sample exhibit tidal features ($7.2\%$).

We assessed the reliability of the tidal feature identification by comparing our primary classifications (performed by Y.Y.) with two independent classifications separately conducted by two authors (K.C. and W.B.) on a subsample of 262 type 1 AGNs with $\log L_\mathrm{[O\,{\footnotesize III}]}>40.8$. The comparison shows that $91\%$ (62/68) and $87\%$ (59/68) of AGN hosts identified as having tidal features by Y.Y. are also classified as having tidal features by K.C. and W.B., respectively. Additionally, $94\%$ (183/194) and $97\%$ (188/194) of AGN hosts without tidal features in Y.Y.'s classification are also identified as having no tidal features by K.C. and W.B., respectively. The outcome of this comparison indicates that the detection of tidal feature is highly consistent across different identifiers, even though it relies on visual inspection, which can be subjective. The high consensus rate over $\sim90\%$ for tidal feature identification is also found in our previous studies on the sample of early-type galaxies \citep{YL2020,Yoon2022,Yoon2024b}. The consensus rate of $\sim90\%$--$95\%$ can introduce an additional potential uncertainty of $\sim0.01$--$0.08$ to the fraction of tidal features presented in Sect. \ref{sec:results}, depending on the number of samples per bin. We note that these uncertainties are smaller than the standard errors defined in this study.
\\

\begin{figure}
\includegraphics[width=\linewidth]{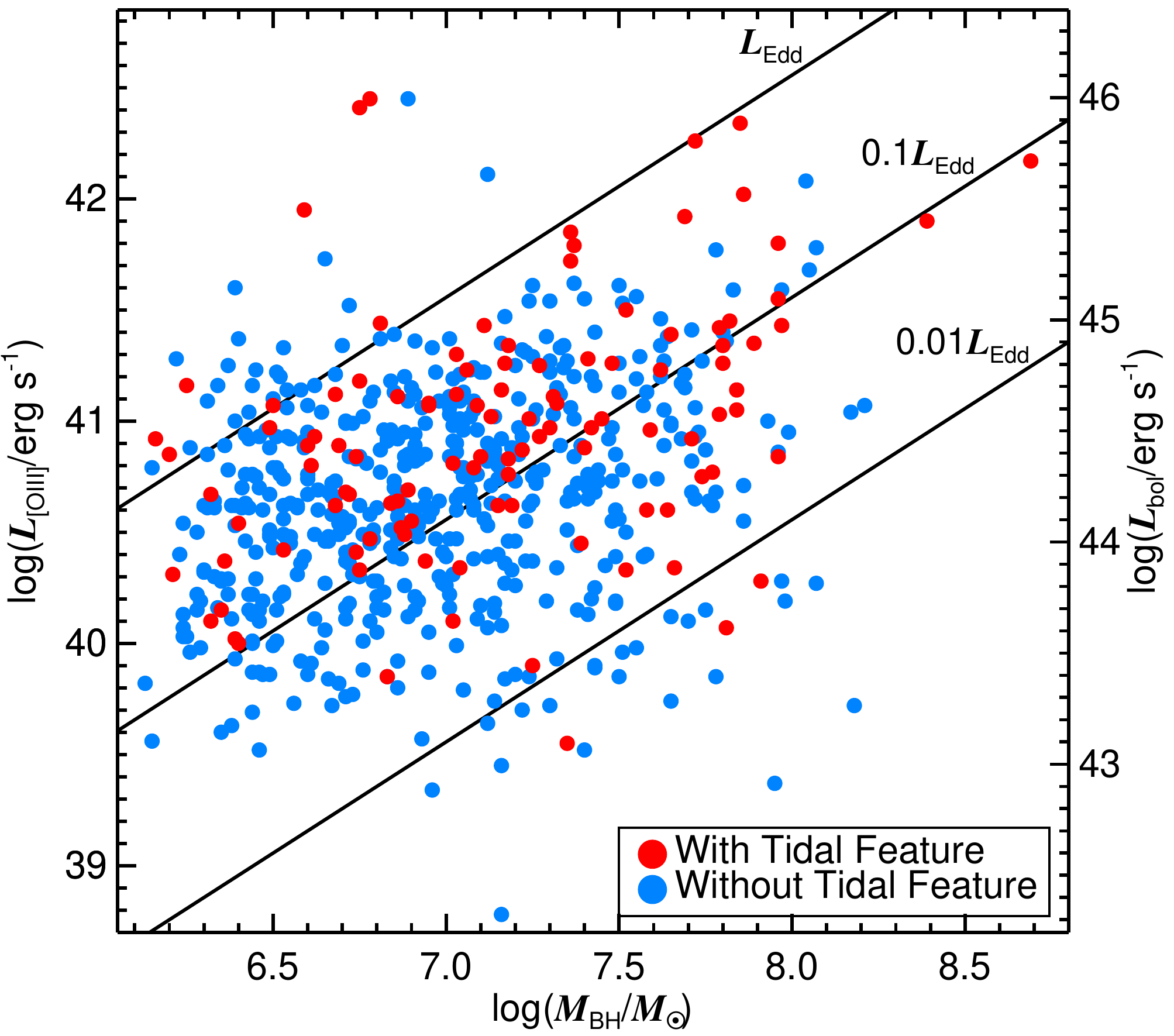}
\centering
\caption{Distribution of type 1 AGNs in the plane of logarithmic values of $M_\mathrm{BH}$ vs. $L_\mathrm{[O\,{\footnotesize III}]}$ (or $L_\mathrm{bol}$). AGNs are categorized according to the presence or absence of tidal features. The solid black lines indicate constant values of $L_\mathrm{Edd}$, $0.1L_\mathrm{Edd}$, and $0.01L_\mathrm{Edd}$.
\label{fig:mo3}}
\end{figure}

\begin{figure*}
\includegraphics[width=\linewidth]{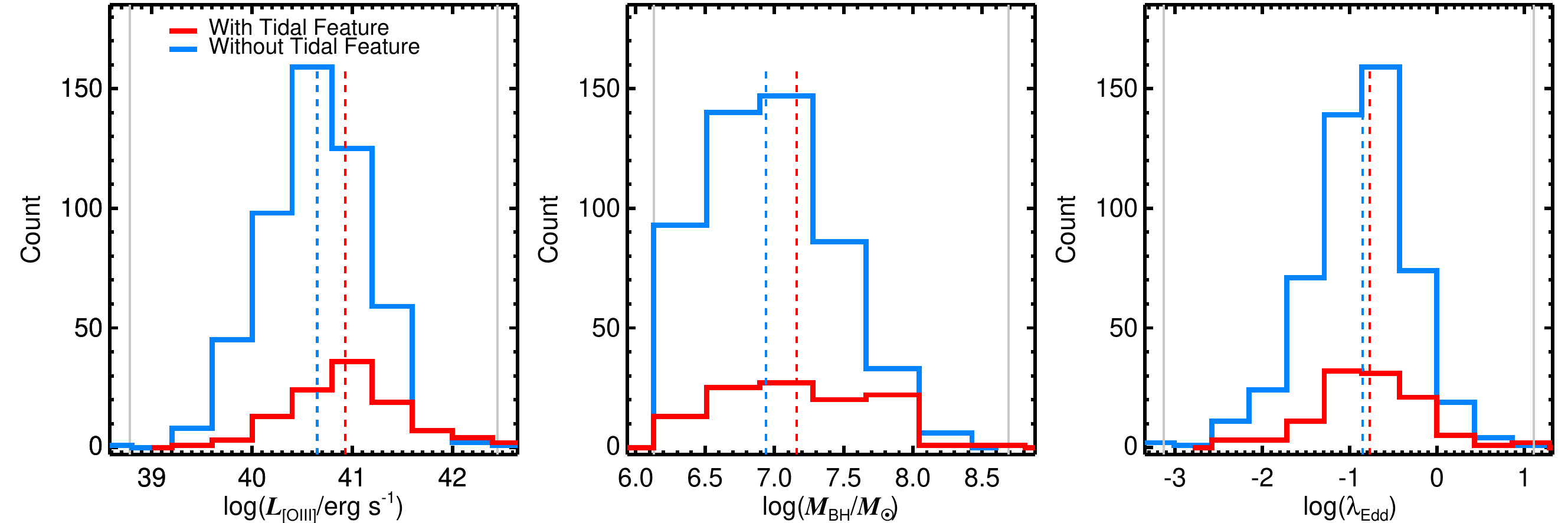}
\centering
\caption{Distributions of $\log L_\mathrm{[O\,{\footnotesize III}]}$, $\log M_\mathrm{BH}$, and $\log\lambda_\mathrm{Edd}$ for AGN hosts with and without tidal features. The vertical dashed lines represent the median value of each parameter for the two AGN categories. The gray vertical lines indicate the full range of each parameter.
\label{fig:dist}}
\end{figure*}

\section{Results}\label{sec:results}
We present the properties of our type 1 AGNs and their distributions in Figs. \ref{fig:mo3} and \ref{fig:dist}. Figure \ref{fig:mo3} displays the distribution of type 1 AGNs in the logarithmic values of $M_\mathrm{BH}$ versus $L_\mathrm{[O\,{\footnotesize III}]}$ (or $L_\mathrm{bol}$) plane, in which AGNs are divided based on the presence or absence of tidal features. Figure \ref{fig:dist} shows histograms representing the distributions of $\log L_\mathrm{[O\,{\footnotesize III}]}$, $\log M_\mathrm{BH}$, and $\log\lambda_\mathrm{Edd}$ for AGN hosts with and without tidal features.

Figure \ref{fig:mo3} shows that AGN hosts with tidal features are spread across the region in the $\log M_\mathrm{BH}$ versus $\log L_\mathrm{[O\,{\footnotesize III}]}$ plane that is populated by AGN hosts without tidal features. However, AGN hosts with tidal features are more likely to lie in the region of higher $L_\mathrm{[O\,{\footnotesize III}]}$ and higher $M_\mathrm{BH}$, compared to those without tidal features.

Figure \ref{fig:dist} demonstrates that type 1 AGN hosts with tidal features have median $L_\mathrm{[O\,{\footnotesize III}]}$ and $M_\mathrm{BH}$ values that are 0.28 and 0.22 dex higher, respectively, than those without tidal features. By conducting a Kolmogorov--Smirnov (KS) test on the two distributions of $\log L_\mathrm{[O\,{\footnotesize III}]}$ for the two AGN categories, we find that the probability ($0\le p\le1$) of the null hypothesis, in which the two distributions stem from the same distribution, is $p=3.2\times10^{-5}$. Similarly, a KS test on the two distributions of $\log M_\mathrm{BH}$ for the two AGN populations yields $p=6.2\times10^{-4}$. Thus, these tests prove that AGN hosts with and without tidal features have significantly different distributions of $L_\mathrm{[O\,{\footnotesize III}]}$ and $M_\mathrm{BH}$. AGN hosts with tidal features have a slightly higher median $\log\lambda_\mathrm{Edd}$, by 0.08 dex, than those without tidal features, but this difference is not statistically significant, as a KS test on the two distributions of $\log\lambda_\mathrm{Edd}$ for the two AGN categories yields $p=0.16$. We reach the same conclusions by deriving $p$-values from Anderson--Darling tests conducted on the two distributions of $\log L_\mathrm{[O\,{\footnotesize III}]}$,  $\log M_\mathrm{BH}$, and $\log\lambda_\mathrm{Edd}$ for the two AGN categories. The tests yield $p<0.001$ for  $\log L_\mathrm{[O\,{\footnotesize III}]}$ and $\log M_\mathrm{BH}$, while for $\log\lambda_\mathrm{Edd}$, the $p$-value is $0.13$.

\begin{figure*}
\includegraphics[width=\linewidth]{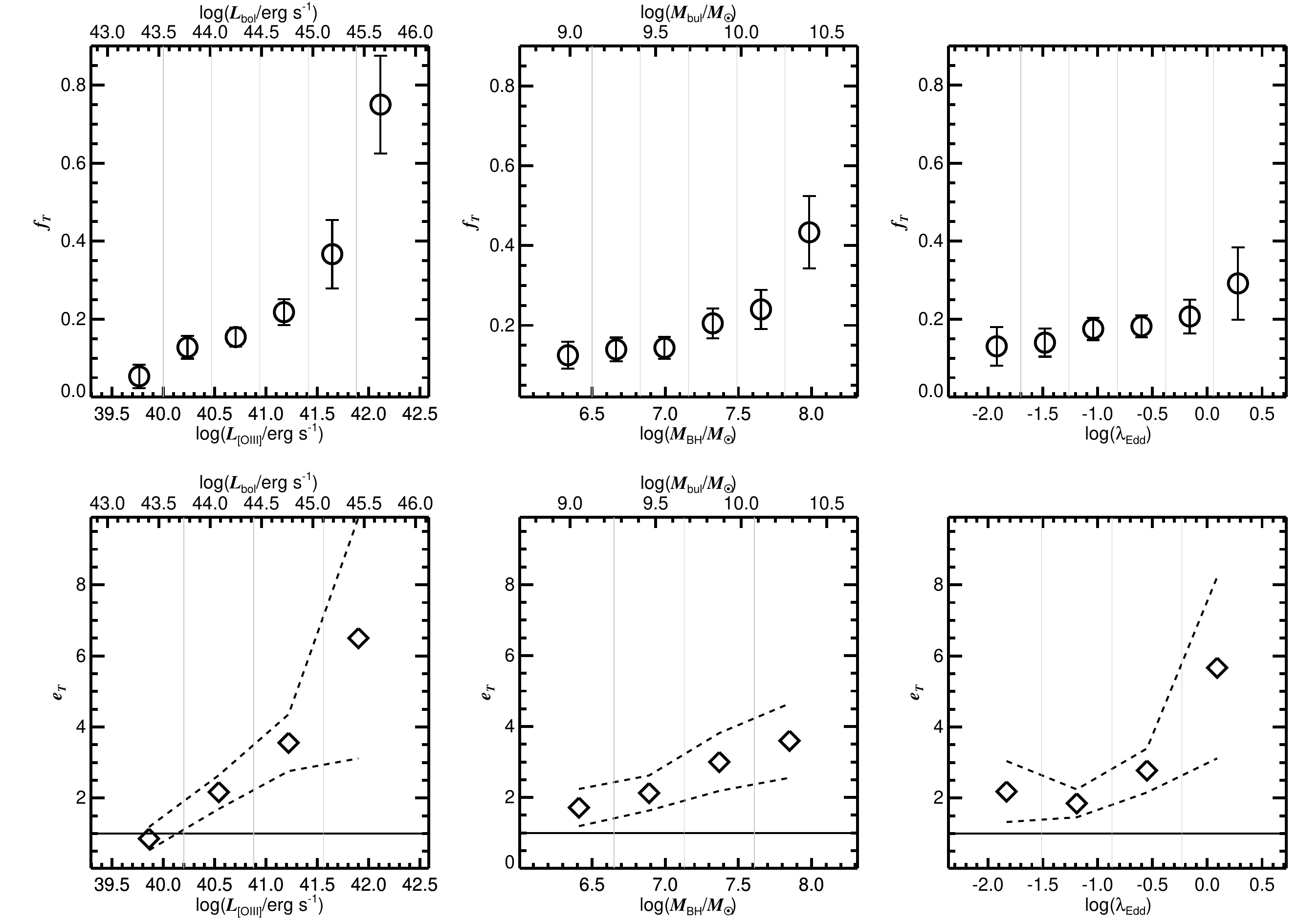}
\centering
\caption{Top panels: Fraction of type 1 AGN hosts with tidal features ($f_T$) as a function of $\log L_\mathrm{[O\,{\footnotesize III}]}$, $\log M_\mathrm{BH}$, and $\log\lambda_\mathrm{Edd}$. The error bar indicates the standard error of the proportion. Bottom panels: Excess of $f_T$ ($e_T$) as a function of $\log L_\mathrm{[O\,{\footnotesize III}]}$, $\log M_\mathrm{BH}$, and $\log\lambda_\mathrm{Edd}$, defined as the ratio of $f_T$ for AGNs to that of the matched inactive control sample. The dashed lines represent the range of the error value, computed through error propagation from the standard errors of the two proportions. The horizontal lines in the bottom panels represent $e_T=1$, which indicates that the $f_T$ of AGNs is identical to that of the inactive control sample. The gray vertical lines in all the panels mark the boundaries of the bins. The bin sizes in the bottom panels are set to be slightly larger (and hence there is a smaller number of bins) than those in the top panels, in order to reduce the larger uncertainties arising from the error propagation of the two proportion errors.
\label{fig:frac}}
\end{figure*}

\begin{figure*}
\includegraphics[scale=0.25]{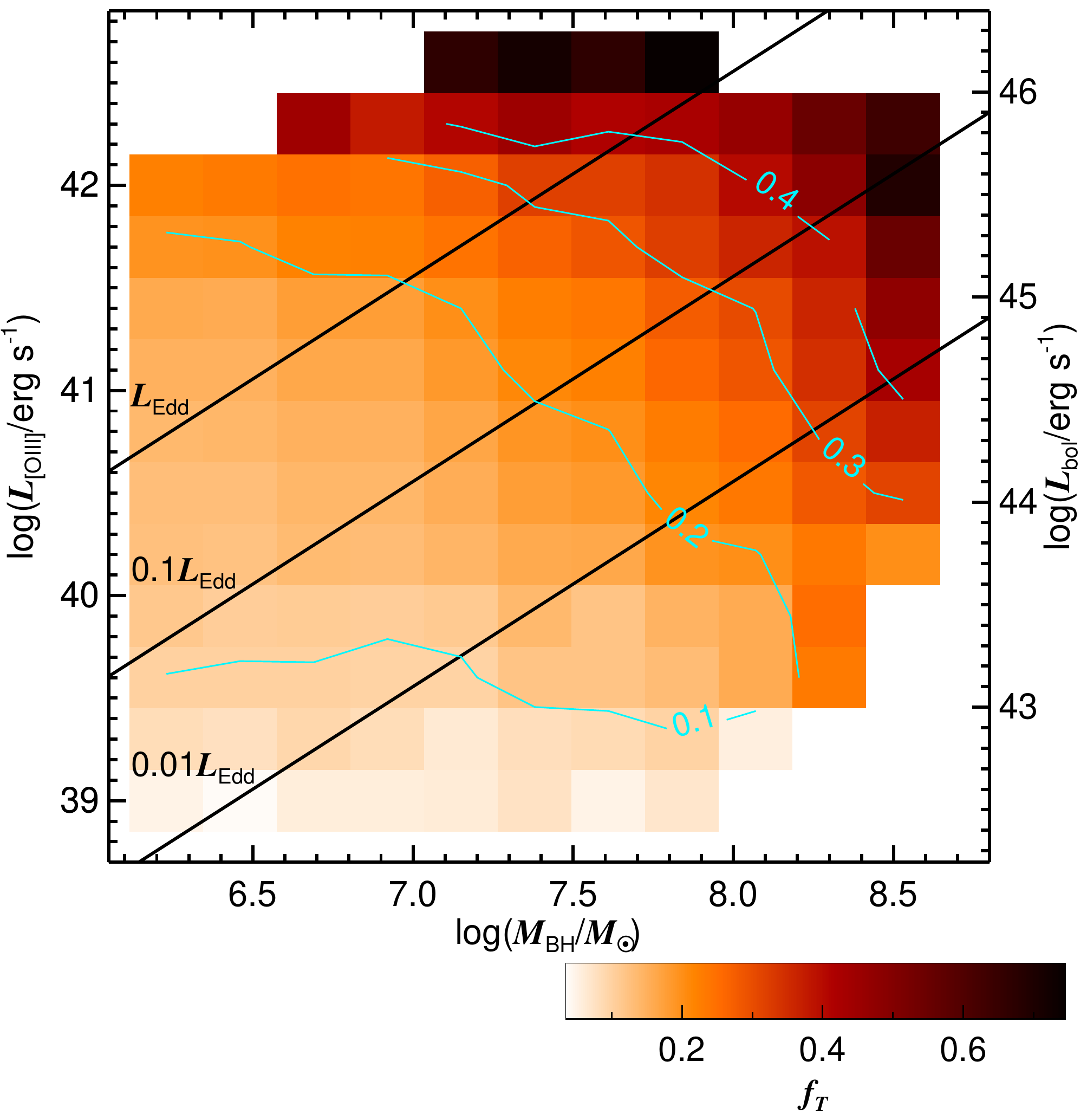}\includegraphics[scale=0.25]{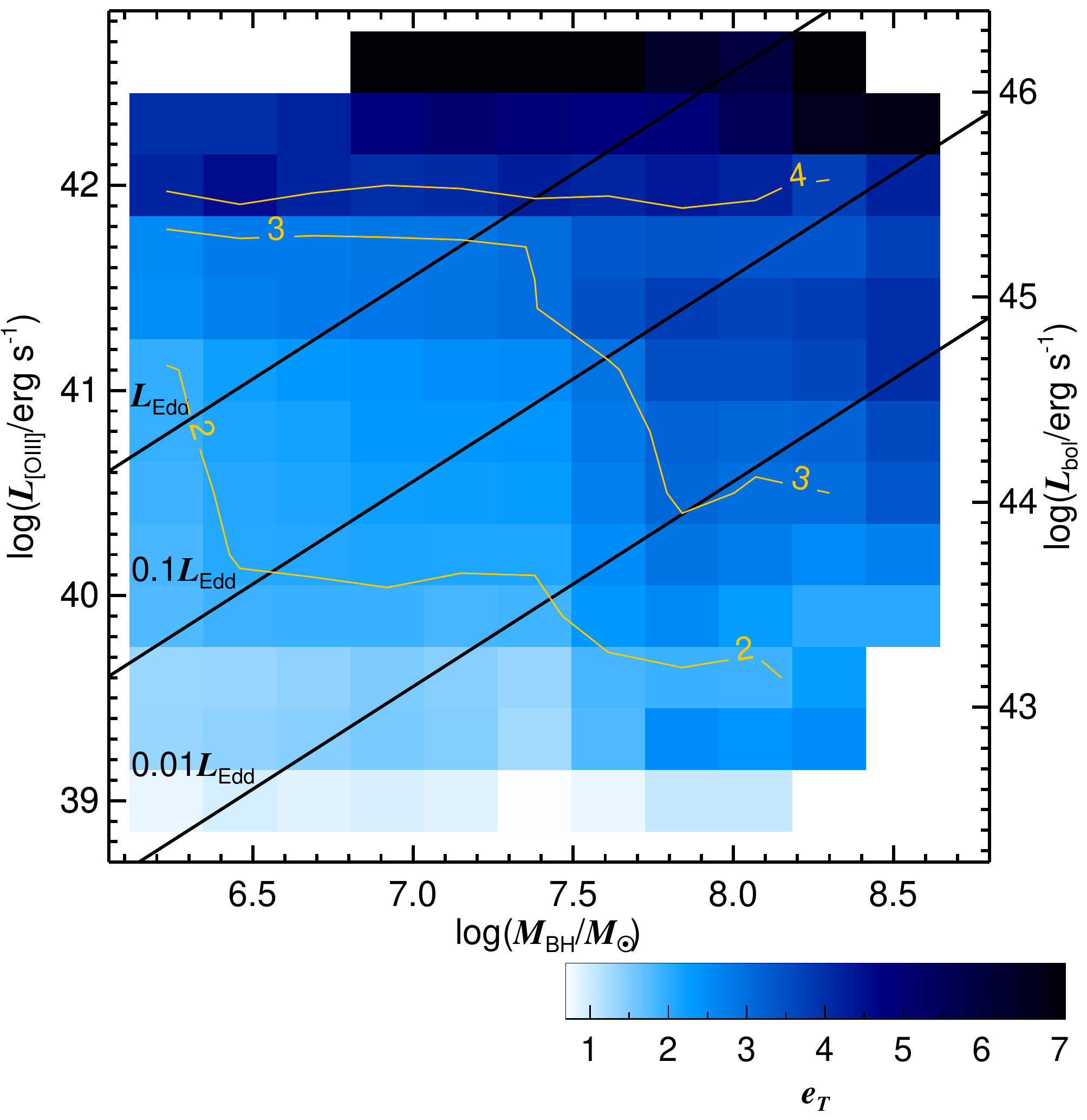}
\centering
\caption{Fraction of type 1 AGN hosts with tidal features ($f_T$) and its excess ($e_T$), defined as the ratio of $f_T$ for AGNs to that of the matched inactive control sample, shown using colors and contours in the $\log M_\mathrm{BH}$ vs. $\log L_\mathrm{[O\,{\footnotesize III}]}$ plane (the left panel is for $f_T$, while the right panel is for $e_T$). See the color bars for the color-coded representations of $f_T$ and $e_T$. The values in the middle of the contour lines denote $f_T$ or $e_T$. To create color maps and contours, we used a grid with block sizes of 0.23 dex and 0.30 dex along the $\log M_\mathrm{BH}$ and $\log L_\mathrm{[O\,{\footnotesize III}]}$ axes, respectively. At each grid point, we calculated $f_T$ using a rectangular bin with side lengths of 1.38 dex and 1.80 dex along the $\log M_\mathrm{BH}$ and $\log L_\mathrm{[O\,{\footnotesize III}]}$ axes, respectively, and computed $e_T$ using a rectangular bin with side lengths of 1.84 dex and 2.40 dex. Using a larger bin size than the grid block size to compute the parameters smooths the color maps and contours, revealing the large-scale trends more clearly. We display only the colored bins for which more than 10 and 25 AGNs are used to compute $f_T$ and $e_T$, respectively.
\label{fig:2d}}
\end{figure*} 

We display the fraction of type 1 AGN hosts with tidal features ($f_T$) as a function of $\log L_\mathrm{[O\,{\footnotesize III}]}$, $\log M_\mathrm{BH}$, and $\log\lambda_\mathrm{Edd}$ in the top panels of Fig. \ref{fig:frac}. The fraction $f_T$ is defined as $N_T/N_\mathrm{AGN}$, where $N_\mathrm{AGN}$ denotes the total number of AGN hosts, and $N_T$ represents the number of AGN hosts with tidal features.\footnote{In this study, the uncertainties of $f_T$ are given by the standard error of the proportion, $\sqrt{f_T(1-f_T)/N_\mathrm{AGN}}$. These errors are nearly identical to those obtained from the bootstrapping method (resampling 2000 times for each bin), differing by no more than $\sim1$--$2\%$.} Figure \ref{fig:frac} demonstrates a rapid increase in $f_T$ with increasing $L_\mathrm{[O\,{\footnotesize III}]}$, such that $f_T$ for the lowest-luminosity AGNs with $\log L_\mathrm{[O\,{\footnotesize III}]}<40.0$ is $0.05\pm0.03$, while for the highest-luminosity AGNs with $\log L_\mathrm{[O\,{\footnotesize III}]}>41.9$, it reaches $0.75\pm0.13$. We conducted a hypothesis test to compare the $f_T$ values in the two extreme $L_\mathrm{[O\,{\footnotesize III}]}$ bins. The null hypothesis states that there is no difference between the $f_T$ values of the highest- and lowest-luminosity AGNs, while the alternative hypothesis is that $f_T$ of the highest-luminosity AGNs is significantly higher than that of the low-luminosity AGNs. Using a one-tailed $Z$-test, we find a $p$-value of $4.7\times10^{-9}$. This implies that $f_T$ of high-luminosity AGNs is significantly higher than that of low-luminosity AGNs.

Figure \ref{fig:frac} also indicates that $f_T$ increases with rising $M_\mathrm{BH}$, as $f_T$ for AGNs with $\log M_\mathrm{BH}<6.5$ is $0.13\pm0.03$, while it is $0.43\pm0.09$ for AGNs with $\log M_\mathrm{BH}>7.8$. For the $f_T$ values in the two extreme $M_\mathrm{BH}$ bins, we performed a one-tailed hypothesis test, in which the null and the alternative hypotheses are of the same form as those described above. This test yields a $p$-value of $1.1\times10^{-4}$, suggesting that $f_T$ of AGNs with high $M_\mathrm{BH}$ is significantly higher than that of AGNs with low $M_\mathrm{BH}$.

The third column in the same figure shows that $f_T$ also increases with $\lambda_\mathrm{Edd}$, though the trend is weaker compared to those with $L_\mathrm{[O\,{\footnotesize III}]}$ and $M_\mathrm{BH}$. Specifically, $f_T$ for AGNs with $\log\lambda_\mathrm{Edd}<-1.7$ is $0.13\pm0.05$, while for AGNs with $\log\lambda_\mathrm{Edd}>0.1$, it is $0.29\pm0.09$. A one-tailed hypothesis test on the $f_T$ values in the two extreme $\lambda_\mathrm{Edd}$ bins, with the null and the alternative hypotheses following the same structure as those described above, yields a $p$-value of 0.05. This suggests that the statistical significance of the trend, where $f_T$ of AGNs with high $\lambda_\mathrm{Edd}$ is higher than that of AGNs with low $\lambda_\mathrm{Edd}$, is relatively weak.

In the left panel of Fig. \ref{fig:2d}, we show $f_T$ in the $\log M_\mathrm{BH}$ versus $\log L_\mathrm{[O\,{\footnotesize III}]}$ plane, in order to illustrate the composite trend of $f_T$ with respect to $L_\mathrm{[O\,{\footnotesize III}]}$, $M_\mathrm{BH}$, and $\lambda_\mathrm{Edd}$.\footnote{We use larger bin sizes than the grid block size in Fig. \ref{fig:2d} to smooth the color maps and contours, enhancing the visibility of large-scale trends. Therefore, for accurate values of $f_T$ or $e_T$ in a specific bin, it is recommended to refer to Fig. \ref{fig:frac}.} This figure also demonstrates that $f_T$ is higher for AGNs with higher luminosities and more massive BHs, while it peaks at the highest-luminosity AGNs with $\log L_\mathrm{[O\,{\footnotesize III}]}\gtrsim42$, showing that the trend is more pronounced with AGN luminosity.

Lastly, we present the excess of $f_T$ ($e_T$) as a function of $\log L_\mathrm{[O\,{\footnotesize III}]}$, $\log M_\mathrm{BH}$, and $\log\lambda_\mathrm{Edd}$ in the bottom panels of Fig. \ref{fig:frac}. The parameter $e_T$ is defined as the ratio of $f_T$ for AGNs in a given bin to that of their matched inactive control galaxies. Figure \ref{fig:frac} shows that $e_T$ increases from $0.9\pm0.3$ to $6.5\pm3.4$ as $L_\mathrm{[O\,{\footnotesize III}]}$ rises, indicating that $f_T$ for low-luminosity AGNs is nearly identical to that of the matched inactive galaxies, whereas high-luminosity AGNs are more than six times as likely to exhibit tidal features as their inactive counterparts. Figure \ref{fig:frac} also displays that $e_T$ increases from $1.7\pm0.5$ to $3.6\pm1.0$ with rising $M_\mathrm{BH}$, although the trend is weaker than that with $L_\mathrm{[O\,{\footnotesize III}]}$. For $\lambda_\mathrm{Edd}$, $e_T$ also increases with $\lambda_\mathrm{Edd}$. At $\log\lambda_\mathrm{Edd}<-0.2$, the rate of increase in $e_T$ is modest (from $\sim2$ to $3$). However, at very high Eddington ratios of $\log\lambda_\mathrm{Edd}>-0.2$, $e_T$ rises sharply to $5.7\pm2.6$. Based on these results, we find that a certain level of excess in $f_T$ exists across all AGNs, except for those with the lowest luminosities.

In the right panel of Fig. \ref{fig:2d}, $e_T$ is shown in the $\log M_\mathrm{BH}$ versus $\log L_\mathrm{[O\,{\footnotesize III}]}$ plane, which displays how $e_T$ varies jointly with $L_\mathrm{[O\,{\footnotesize III}]}$, $M_\mathrm{BH}$, and $\lambda_\mathrm{Edd}$. The figure illustrates that $e_T$ primarily increases with AGN luminosity, with a weaker trend suggesting that AGNs with more massive BHs also exhibit higher $e_T$. The highest-luminosity AGNs with $\log L_\mathrm{[O\,{\footnotesize III}]}\gtrsim42$ exhibit high values of $e_T\gtrsim4$ across all $M_\mathrm{BH}$, including those with low $M_\mathrm{BH}$, which correspond to AGNs with very high $\lambda_\mathrm{Edd}$. This explains the sharp increase in $e_T$ at very high $\lambda_\mathrm{Edd}$ shown in Fig. \ref{fig:frac}. 
\\

\section{Discussion}\label{sec:discuss}

\subsection{Connection between AGNs and galaxy mergers}\label{sec:agnmerger}

Our key result is that $f_T$ and $e_T$ are higher for more luminous AGNs, with $f_T$ and $e_T$ reaching $0.75$ and $\sim6.5$, respectively, for the highest-luminosity AGNs with $\log L_\mathrm{[O\,{\footnotesize III}]}\gtrsim42$. If $f_T$ for AGNs can be interpreted as the fraction of AGNs triggered by mergers, this provides direct observational evidence that galaxy mergers are the predominant triggering mechanism for such high-luminosity AGNs. In addition, the progenitor galaxies involved in mergers that trigger high-luminosity AGNs are likely to be rich in gas \citep{Byrne2023,Avirett2024}, as sustaining high-luminosity AGN radiation requires substantial gas accretion onto the BH \citep{Byrne2023}.

The fraction $f_T$ of the lowest-luminosity AGNs with $\log L_\mathrm{[O\,{\footnotesize III}]}\lesssim40$ is very low (0.05) and does not significantly differ from that of their inactive counterparts. This suggests that such low-luminosity AGNs are unlikely to be triggered by galaxy mergers; instead, other mechanisms mentioned in Sect. \ref{sec:intro} may be responsible for their triggering. For instance, according to \citet{Hirschmann2012}, disk instabilities can serve as a trigger mechanism for low- to moderate-luminosity AGNs with $\log L_\mathrm{bol}\lesssim45$ (equivalent to $\log L_\mathrm{[O\,{\footnotesize III}]}\lesssim41.5$) in the low-redshift Universe. Given that bar-driven fueling of central SMBHs has been observed in Seyfert galaxies \citep{Crenshaw2003,Ohta2007}, which host moderate-luminosity AGNs, bars can also trigger low- to moderate-luminosity AGNs with $\log L_\mathrm{bol}\lesssim45$. Ram pressure in cluster environments can trigger low-luminosity AGNs, as supported by the sample in \citet{Poggianti2017}, in which AGN hosts experiencing ram pressure typically have $\log L_\mathrm{[O\,{\footnotesize III}]}\lesssim41$.

According to \citet{Durret2021}, who studied 40 galaxy cluster samples, $\sim10\%$ of galaxies in clusters are experiencing ram pressure. This fraction can therefore be considered an upper limit to the population of ram-pressure-triggered AGNs in rich environments such as clusters, under the crude assumption that ram pressure is a highly effective AGN triggering mechanism. We matched our AGN sample to the large redshift-survey group catalog of \citet{Tempel2014}, which is constructed from SDSS DR10. Among the 538 matched AGNs, we find that $58\%$ of AGNs reside in group or cluster environments, while $13\%$ are located in richer environments with $\geq10$ spectroscopic members, a typical threshold in observational work used to distinguish environments denser than poor clusters or rich groups \citep{Miller2005,Porter2005}. Combining these figures, we estimate that roughly $\lesssim1$--$5\%$ of our AGN sample may be undergoing ram pressure. Future studies of AGN environments using large survey data will provide a more reliable estimate of the fraction of ram-pressure-triggered AGNs.

The other AGNs, except those in the highest- and lowest-luminosity bins, do not exhibit $f_T$ higher than 0.4 but show the excess in $f_T$ compared to their inactive counterparts. This implies that galaxy mergers contribute to triggering of these AGNs, although mergers are not the dominant triggering mechanism. Low- to moderate luminosity AGN populations are far more numerous than high-luminosity AGNs, which are primarily triggered by mergers. Therefore, it can be said that most AGNs are generally triggered by mechanisms other than galaxy mergers, as suggested by several studies \citep{Treister2012,Steinborn2018,Man2019,McApine2020,Hernandez2023}.

Since more massive galaxies tend to host more massive SMBH \citep{Kormendy2013}, our finding that $f_T$ is higher for AGNs with higher $M_\mathrm{BH}$ is consistent with the previous studies showing that more massive galaxies are more likely to exhibit tidal features, suggesting a higher likelihood of recent mergers in more massive galaxies \citep{Hong2015,YL2020,Yoon2024a}. However, the rate of increase in $f_T$ with $M_\mathrm{BH}$ is higher for AGNs than for their inactive counterparts, as $e_T$ is higher at higher $M_\mathrm{BH}$. Although this trend is not very strong, it implies that galaxy mergers play a more crucial role in activating more massive SMBHs, as shown in the simulation by \citet{Hopkins2014}. This trend may be related to the fact that lower-mass galaxies have a higher gas fraction \citep{Hopkins2009,Masters2012}. In this case, due to the abundant gas in lower-mass galaxies, mild mechanisms, other than mergers, can sufficiently induce gas inflow onto their SMBHs. In contrast, a stronger event, such as galaxy mergers, is more necessary to activate massive SMBHs in more massive galaxies due to their lower gas fraction, leading to increased $e_T$ with increasing $M_\mathrm{BH}$.

According to \citet{Pierce2023}, the surface brightness depth of images is an important factor in studying merger features. Thus, our results may be affected by image depth. For example, a lot of very faint tidal features, which are only detectable through far deeper images, may be hidden around low-luminosity AGN hosts, meaning that the actual $f_T$ for those AGNs could be much higher. Even if that is the case, we can still conclude that tidal features prominent enough to be detected in this study are more frequently found around AGNs with higher luminosities and more massive $M_\mathrm{BH}$, which means that these AGNs are more likely associated with more significant or recent mergers than their low-luminosity or less massive counterparts. Moreover, the excess $e_T$ is unlikely to be significantly influenced by image depth, as demonstrated by \citet{Ellison2019}.
\\

\subsection{Near- and super-Eddington AGNs and the uncertainties in Eddington ratios}\label{sec:eddington}

In our AGN sample, $5\%$ have $\lambda_\mathrm{Edd}$ higher than 1. The large uncertainty associated with observed Eddington ratios may result in AGNs with intrinsically high Eddington ratios being identified as near- or super-Eddington AGNs. This uncertainty arises from several factors. Specifically, the conversion from $L_\mathrm{[O\,{\footnotesize III}]}$ to $L_\mathrm{bol}$ may involve an uncertainty larger than $\sim0.4$ dex \citep{Heckman2004}. Additionally, BH mass measurements based on single-epoch spectra are known to have uncertainties of a factor of 3 \citep{Greene2005}. Combined, these factors can result in uncertainties of $\sim0.7$ dex in the Eddington ratios calculated in this study.

Another source of uncertainty that can lead to high observed Eddington ratios is the presence of narrow-line Seyfert 1 galaxies, whose broad emission lines have FWHMs of less than $\sim2000$ km s$^{-1}$. According to \citet{Marconi2008}, $M_\mathrm{BH}$ values for this type of AGN can be underestimated by more than $\sim0.5$ dex, potentially contributing to the apparent excess of super-Eddington AGNs. However, excluding AGNs with FWHMs of broad H$\alpha$ lines below 2000 km s$^{-1}$ (which account for $9\%$ of our AGN sample) does not change our results, suggesting that the bias introduced by this effect is negligible. 

However, it is also possible that some of the observed super-Eddington AGNs are genuinely accreting at such high rates. The structure of real AGNs often deviates significantly from the assumption of spherical symmetry used in calculating Eddington ratios, allowing for super-Eddington accretion to be physically possible in certain cases \citep{Du2015}. Taken together, while measurement uncertainties and potential narrow-line Seyfert 1 galaxies can inflate Eddington ratios, departures from spherical symmetry suggest that some objects are genuinely near- or super-Eddington. Accordingly, the Eddington ratios presented here should be interpreted with caution.
\\

\subsection{Comparison with previous studies}\label{sec:agnmerger}

As mentioned earlier in Sect. \ref{sec:intro}, the excess of the merger fraction in AGN hosts is a controversial issue, but we find the direct evidence of a strong AGN--merger connection, particularly for very luminous AGNs. This demonstrates the importance of including AGNs with a wide range of luminosities in the sample to develop a comprehensive understanding of the AGN--merger connection. Redshift ranges of AGN samples can also be an important factor contributing to the different results, as most studies that found no evidence of a connection between mergers and AGNs use samples at higher redshifts ($0.5\lesssim z \lesssim3.0$; \citealt{Gabor2009,Cisternas2011,Kocevski2012,Mechtley2016,Villforth2014,Villforth2017,Marian2019,Shah2020}), while those suggesting the AGN--merger connection, including this study, typically focus on lower redshifts ($z\lesssim0.2$; \citealt{Carpineti2012,Cotini2013,Ellison2013,Hong2015,Marian2020,Araujo2023,Hernandez2023,Li2023,Comerford2024}). The difference in results due to the redshift range of the sample may stem from the fact that mergers play an increasingly important role in triggering AGNs as the Universe evolves, even though the abundance of merging systems decreases with decreasing redshift \citep{McApine2020}. In addition, the difficulty in detecting merger or disturbance features in high-redshift galaxies, due to cosmological surface brightness dimming and small angular sizes, may have influenced studies of high-redshift AGNs, as cosmological dimming has been identified as one of the most important factors affecting the study of merger features, according to \citet{Pierce2023}.

Comparing our results with previous studies supporting the AGN--merger connection allows us to broaden our understanding of the general AGN population, as our analysis here is limited to type 1 AGNs selected through spectroscopy. Using type 2 AGNs with $40.5\lesssim\log L_\mathrm{[O\,{\footnotesize III}]}\lesssim42.5$, which overlaps with the luminosity range of our sample, \citet{Pierce2022} and \citet{Pierce2023} show that type 2 AGN hosts exhibit a substantially enhanced rate of morphological disturbance, by a factor of $\sim2$--$3$, compared with a matched control sample. Furthermore, the disturbance rate increases from $30\%$ to $70\%$ across the full luminosity range, accompanied by an increase in the enhancement factor relative to the control sample from $1$ to $3$. These findings are highly consistent with our results. \citet{Urbano2019} suggest that low-luminosity type 2 AGNs with $\log L_\mathrm{[O\,{\footnotesize III}]}\lesssim41.9$ exhibit a low fraction of disturbed morphology ($6\%$) compared to high-luminosity type 2 AGNs with $\log L_\mathrm{[O\,{\footnotesize III}]}\gtrsim41.9$, which show a fraction of $34\%$. In addition, several previous studies support the AGN--merger connection for type 2 AGNs \citep{Carpineti2012,Araujo2023,Li2023}.

Several studies based on type 1 AGNs report results consistent with ours. For instance, \citet{Urrutia2008} found that over $\sim80\%$ of intrinsically bright type 1 AGNs exhibit strong evidence of recent or ongoing mergers, based on dust-reddened quasars. \citet{Tang2023} reported a correlation between $L_\mathrm{bol}$ and host asymmetry in type 1 AGNs, which strengthens significantly at the highest luminosities of $\log L_\mathrm{bol}>45$. Similarly, \citet{Hernandez2023} found that both type 1 and type 2 AGNs show a higher incidence of tidal features than a non-AGN control sample. Given that the difference between type 1 and type 2 AGNs arises from the viewing angle, according to the AGN unified model, it is reasonable that the AGN--merger connection and its luminosity-dependent trend are similarly observed in both types.

Multiple previous studies based on AGNs selected via X-ray or infrared observations are also consistent with our results, supporting the AGN--merger connection. For example, using an AGN sample selected from X-ray, infrared, and spectroscopic surveys, \citet{Treister2012} find a strong correlation between AGN luminosity and the fraction of host galaxies undergoing mergers. Specifically, the fraction increases from $\sim4\%$ at $\log L_\mathrm{bol}\sim43$ to $\sim80\%$ at $\log L_\mathrm{bol}\sim46$, which is very similar to our result. Other studies also support the AGN--merger connection based on AGNs selected via X-ray or infrared (e.g., \citealt{Cotini2013,Li2023}). Thus, these results, combined with ours, imply that the AGN--merger connection, particularly for high-luminosity AGNs, can be identified across AGN samples selected through various wavelength windows.
\\

\section{Summary}\label{sec:summary}

Using a large sample of 614 type 1 AGNs at $z<0.07$, we explored whether galaxy mergers are a significant triggering mechanism for AGNs and identified the AGN properties most associated with mergers, aiming to gain a clearer understanding of the AGN--galaxy merger connection. For the quantitative comparison, we used a control sample of inactive galaxies that are matched to AGNs by $M_\mathrm{BH}$ and redshift. We identified tidal features, which are direct evidence of recent mergers, through visual inspection of DESI Legacy Survey images with a surface brightness depth of $\sim27$ mag arcsec$^{-2}$, a level that previous studies have shown to be sufficient for the detection of tidal features \citep{Kaviraj2010,Schawinski2010,Hong2015}. We focused on two parameters: the fraction of type 1 AGN hosts with tidal features ($f_T$) and its excess ($e_T$), which is defined as the ratio of $f_T$ for AGNs to that of the matched inactive control sample.

We discover that $f_T$ is higher for AGNs with higher luminosities and (to a lesser extent) more massive BHs. Specifically, $f_T$ rapidly increases from $0.05\pm0.03$ to $0.75\pm0.13$ as $L_\mathrm{[O\,{\footnotesize III}]}$ rises in the range $10^{39.5}\lesssim L_\mathrm{[O\,{\footnotesize III}]}/(\mathrm{erg\,s}^{-1}) \lesssim10^{42.5}$, while $f_T$ increases from $0.13\pm0.03$ to $0.43\pm0.09$ as $M_\mathrm{BH}$ increases in the range $10^{6.0}\lesssim M_\mathrm{BH}/M_{\odot}\lesssim10^{8.5}$. The fraction $f_T$ also increases with $\lambda_\mathrm{Edd}$, though the trend is less significant compared to those with $L_\mathrm{[O\,{\footnotesize III}]}$ and $M_\mathrm{BH}$.

We also find that $e_T$ primarily increases with AGN luminosity, with a weaker trend indicating that $e_T$ is higher for AGNs with more massive BHs. Specifically, $e_T$ increases from $0.9\pm0.3$ to $6.5\pm3.4$ as $L_\mathrm{[O\,{\footnotesize III}]}$ rises, where as $e_T$ increases from $1.7\pm0.5$ to $3.6\pm1.0$ with increasing $M_\mathrm{BH}$.

Our findings offer direct observational evidence that, in the local Universe, galaxy mergers are the predominant triggering mechanism for high-luminosity AGNs, whereas they play a lesser role in the triggering of lower-luminosity AGNs. In addition, our result suggest that a strong event, such as galaxy mergers, is more necessary to activate massive SMBHs in more massive galaxies because of their lower gas fractions. 

With deeper images from future large surveys, we will classify tidal features into several types and explore how different types of mergers relate to AGN properties, as each tidal feature type reflects mergers with distinct natures and origins \citep{Yoon2024b}.
\\

\section{Data availability}
Table \ref{table} is available at the CDS via \url{https://cdsarc.cds.unistra.fr/viz-bin/cat/J/A+A/vol/page}.

\begin{acknowledgements}
This research was supported by Kyungpook National University Research Fund, 2025.
This research was supported by the National Research Foundation of Korea (NRF) grant funded by the Korea government (MSIT) (RS-2025-16064514).
Y. K. was supported by the faculty research fund of Sejong University in 2025 and the National Research Foundation of Korea (NRF) grant funded by the Korean government (MSIT) (No. 2021R1C1C2091550).
D.K. acknowledges the support by the National Research Foundation of Korea (NRF) grant (No. 2021R1C1C1013580) funded by the Korean government (MSIT).
W.B. was supported by the Korea Astronomy and Space Science Institute under the R\&D program (Project No. 2025-1-831-00) supervised by the Ministry of Science and ICT.
\end{acknowledgements}

\end{document}